\documentclass[preprint,12pt]{elsarticle}

\usepackage{amssymb}
\usepackage{hyperref}
\usepackage{xcolor}
\usepackage[symbol*]{footmisc}
\setfnsymbol{wiley}

\journal{SoftwareX}

\begin{document}

\begin{frontmatter}



\title{\texttt{Galclaim}: A tool to identify host galaxy of astrophysical transient sources}


\author{J.-G.~Ducoin$^{a,}$\footnote{Corresponding author: ducoin@iap.fr (J.-G. Ducoin).}}

\affiliation{organization={Sorbonne Université, CNRS, UMR 7095, Institut d’Astrophysique de Paris},
            addressline={98 bis bd Arago}, 
            city={Paris},
            postcode={75014},
            country={France}}

\begin{abstract}
The \texttt{Galclaim} software is designed to identify association between astrophysical transient sources and host galaxy by computing the probability of chance alignment. It is distributed as an open source Python software. It is already used to identify, confirm or reject host galaxy candidates of GRBs and to validate or invalidate transient candidates in astrophysical observations. Such tools are also very useful to characterise archived transient candidates in large sky survey telescopes.

\end{abstract}




\end{frontmatter}


\section*{Metadata}
\label{}

\begin{table}[!ht]
\begin{tabular}{|l|p{6.5cm}|p{6.5cm}|}
\hline
\textbf{Nr.} & \textbf{Code metadata description} & \textbf{Please fill in this column} \\
\hline
C1 & Current code version & v1.0 \\
\hline
C2 & Permanent link to code/repository used for this code version & \url{https://github.com/jgducoin/galclaim} \\
\hline
C3  & Permanent link to Reproducible Capsule & None\\
\hline
C4 & Legal Code License   & GNU General Public License 3 \\
\hline
C5 & Code versioning system used & git and gitlab \\
\hline
C6 & Software code languages, tools, and services used & python \\
\hline
C7 & Compilation requirements, operating environments \& dependencies & numpy, healpy, astropy, matplotlib\\
\hline
C8 & If available Link to developer documentation/manual & None\\
\hline
C9 & Support email for questions & ducoin@iap.fr\\
\hline
\end{tabular}
\caption{Code metadata}
\label{codeMetadata} 
\end{table}

\section{Motivation and significance}
\label{section:motivation}

Identifying galaxies that host transient phenomenons is useful to investigate their environment, provide clues about their formation and the astrophysical condition necessary for their formation (e.g. \citep{Bloom2002,Berger2010,Chrimes2018,Fong2022}). Inferring host candidates is also a fruitful strategy to identify, reject and classify the transients in observations. Such approaches (along with lightcurve classification) are very useful for large sky survey telescopes producing a huge number of transient candidates every night (e.g. \citep{ZTFarchive,Ivezic2019}). But, without spectroscopic observation to provide a clear indication, the association of extra-galactic transient sources with their host galaxy is often a complex task, especially when their localisation is greater than $\sim 1$ arcsec (e.g. \citep{Berger2010}). In order to address this issue, people rely on probability of chance alignment computation, introduced by \cite{Bloom2002}, to statistically provide host candidate (e.g. \citep{Berger2010, Chrimes2018}).

The \texttt{Galclaim} software addresses these concerns providing an open source Python software dedicated to identify association between astrophysical transient sources and host galaxy by estimating the chance alignment between a given transient sky localisation and galaxies identified in astronomical surveys. The code lives in a dedicated git repertory\footnote{\url{https://github.com/jgducoin/galclaim}, \url{https://gitlab.in2p3.fr/ducoin/galclaim}}.

\section{Software description}
\label{section:softwaredescription}

The \texttt{Galclaim} software is distributed as an open source Python software relying on open source and standard tools for data science, numerics, astrophysics and plotting such as numpy, healpy, astropy, astroquery and matplotlib \cite{numpy,healpy,astropy,astroquery,matplotlib}.

The \texttt{Galclaim} software is dedicated to identify association between astrophysical transient sources and host galaxy. This association is made by estimating the chance alignment between a given transient sky localisation and nearby galaxies using the formalism firstly introduced by \cite{Bloom2002} and widely used since in the transient sky community (e.g. \citep{Berger2010, Chrimes2018, Fong2022}). In this formalism, the probability of chance alignment for a given transient and a given galaxy $i$ is expressed as:
\begin{equation}
   P_{i} = 1 - e^{-\pi r^{2}_{i} \sigma(\leq m_{i})}
\label{equation:PvalAsso}
\end{equation}

Where $r_{i}$ is the angular distance between the transient localisation and the galaxy center, $\sigma(\leq m_{i})$ is the number of galaxies per arcsecond square having a magnitude below $m_{i}$ (magnitude of the galaxy $i$). This approach is typically used for transients localised with an error of up to few arcseconds (see Section 5.6.2 of \citep{thesisducoin}). The \texttt{Galclaim} software is relying on sky survey catalogs to crossmatch the transient localisation with known astrophysical sources. The current version of the code uses the Pan-STARRS catalog \cite{PScat}, the Hubble Source Catalog (HSC) \cite{HSCcat}, the AllWISE catalog \cite{AllWISE} and the GLADE catalog \cite{GLADE}. The Pan-STARRS and HSC catalogs provide a good resolution and a relatively deep photometry, two essential properties for the chance alignment estimation, but none of them is all-sky. The Pan-STARRS catalog is limited to $-30$ deg in declination and the HSC one is a visit-based discontinuous catalog. For the sake of completeness of usable sky position, we furthermore implemented the use of the AllWISE infrared catalog \cite{AllWISE} which has the advantage of being all-sky but with worse resolution and depth. Finally, we implemented the GLADE catalog, which is dedicated for multi-messenger searches, as it provide the redshift for nearby galaxies (with a reasonable completeness up to $\sim$ 91Mpc \cite{GLADE}).

In the absence of redshift, the first step is often to identify large galaxies near the transient localisation. For this reason, before any computation, \texttt{Galclaim} pre-check for nearby bright galaxies using the RC3 catalog \cite{RC3} and a 30 arcseconds radius. When a nearby galaxy is found, a warning is raised to the user and the properties of the galaxy are saved in a dedicated output file.

As we are interested on galaxies only, we should identify galaxies in the used catalogs. In the HSC we simply used the extended flag provided. In the Pan-STARRS catalog, we discard stars from galaxies applying the color criteria $(i_{mag,PSF}-i_{mag,Kron})\geqslant 0.05$ up to a magnitude of 21 as proposed by \cite{PScat}. We flag as unknown classification object with magnitude higher than 21 or objects without $i_{mag,Kron}$. For the AllWISE catalog, we used the color criteria  $W1_{mpro} - J_{2MASS} < -1.7$ (in the range of $12<W1_{mpro}<15$) proposed by \cite{Kovacs2015} to select galaxies. We flag as unknown objects for which this criteria is not computable. For both Pan-STARRS and AllWISE catalogs, if the photometry is not sufficient or available for a given object to apply these color criteria, we decide to keep such unknown objects (i.e. treat them as galaxies) in the following computation as it will lead to a slightly over estimation of the galaxy density, i.e. penalise the alignment chance probability and hence harden any significant association. In the case of an association with such unknown object, we leave it up to the user to further investigate if the object is indeed a galaxy.

\begin{figure}
\begin{center}
\includegraphics[width=1\columnwidth]{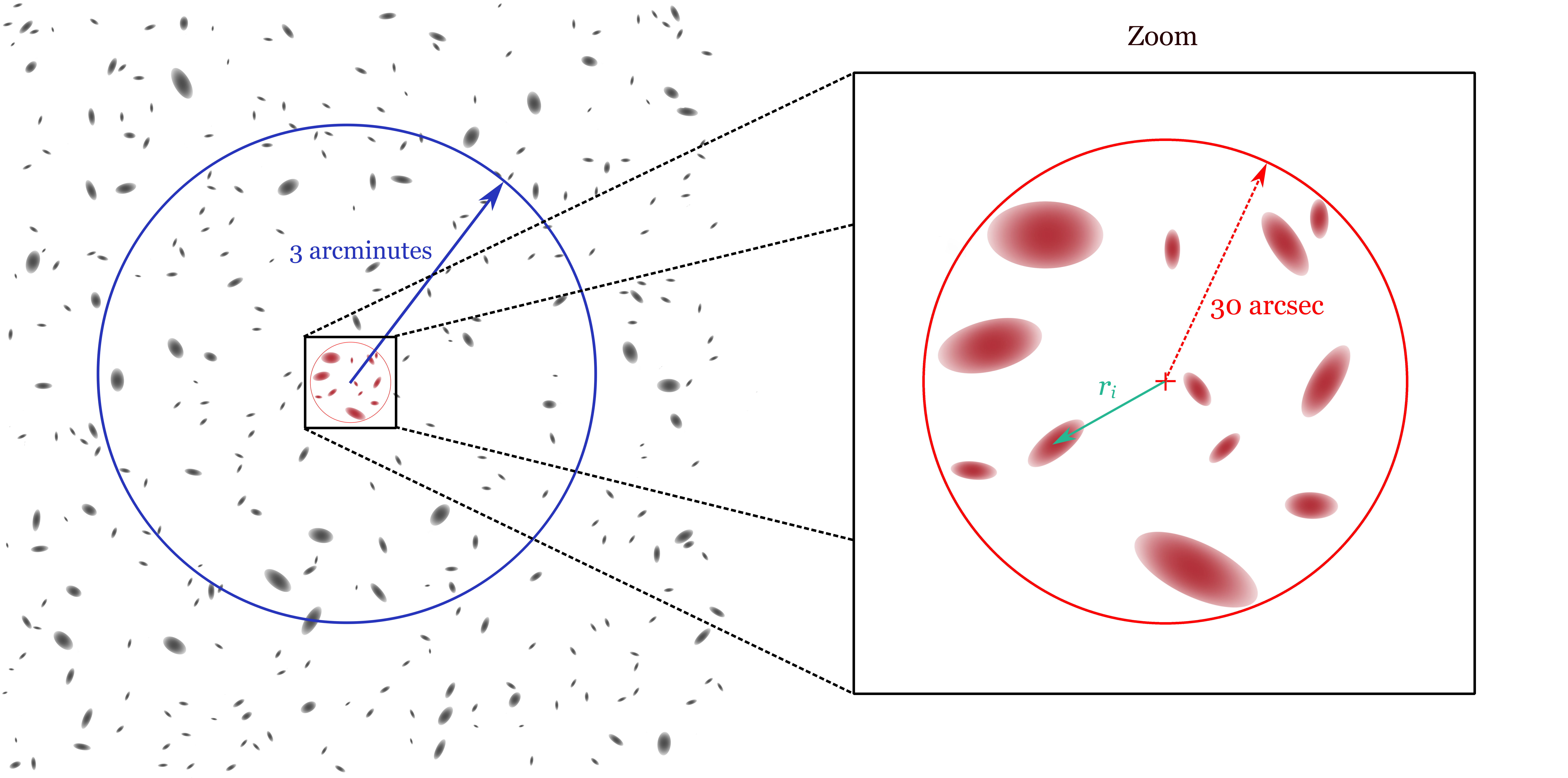}
\caption{Illustration of the probability of chance alignment computed with Equation \ref{equation:PvalAsso}. The blue circle illustrate the 30 arcseconds radius used to compute the galaxy density. The red circle illustrate the 30 arcseconds radius used to consider a galaxy as a host galaxy candidate. The probability of chance alignment is computed for all galaxies in this red circle. The blue and red circle are centered on the transient localisation. The green arrow illustrates the computation of the angular distance between the transient localisation and a given galaxy center.}
\label{fig:illustration}
\end{center}
\end{figure}

To compute the $\sigma$ parameter the \texttt{Galclaim} software follow the principle proposed by \cite{Chrimes2018} which is based on a local estimation of the galaxy density in a given catalog. This allows to take into account the galaxy clustering, as opposed to $\sigma$ estimation based on a whole catalog or using deep optical galaxy surveys as in \cite{Bloom2002,Berger2010}. In practice, as illustrated by Figure \ref{fig:illustration}, we first fetch all galaxies within a 30 arcseconds radius centered on the transient localisation center and consider all of them as host galaxy candidates. We then estimate the galaxy density fetching all galaxies within a 3 arcminutes radius from the transient position center where we remove the host galaxy candidates not to bias the estimation of the local galaxy density with the host itself. For host galaxy candidate $i$ and for each photometric band in the catalog we can then compute the $\sigma(\leq m_{i})$ counting the number of galaxies, within the shell from 30 arcseconds to 3 arcminutes, with magnitude $\leq m_{i}$. We note that, for the HSC, we follow the same procedure but as the all-sky coverage is sparse and discontinuous we modify the way galaxies used to compute the galaxy density are retrieved. We identify in each band of the catalog in which image lies the GRB localisation and use all the galaxies within this image. Knowing the field of view of the catalog images ($202\times 202$ arcseconds square) we have a cross-matched area which is comparable with the 3 arcminutes radius area used for Pan-STARRS, AllWISE and GLADE. 

With $\sigma(\leq m_{i})$  we can compute the $P_{i}$ as described in equation \ref{equation:PvalAsso} for each host galaxy candidates, each catalogs and each photometric band. One can then discuss the association with a given host galaxy candidate looking at the minimum of the computed $P_{i}$ for this galaxy as the result of the probability of chance alignment. While using the GLADE catalog, one can enable the check for compatible redshift providing a redshift range. \texttt{Galclaim} then compute for each host galaxy candidates if the redshift provided in the GLADE catalog is compatible with the provided redshift range and provides this information in the outputted host galaxy candidate properties.

\subsection{Chance alignment threshold}
\label{subsection:threshold}

In order to claim for an association, one needs to define a threshold in $P_{i}$ below which an association can be considered as reliable. While the typical value of the threshold is considered to be around $P_{i} = 0.01$, this threshold is not fixed in \texttt{Galclaim} as different value can be considered depending the transients studied. As the computation of $P_{i}$ include the angular distance between the transient and the galaxy $r_{i}$, different astrophysical sources will lead to different typical value of $P_{i}$.

\subsection{Software Architecture}
\label{subsection:architecture}

The \texttt{Galclaim} software has a very simple structure separated in several python scripts, parsed so that the user can launch the code using a command line in a console. The main dependencies and installation instructions, as well as the usage instructions are described in a usual README file. We provide an example of transient source file format that the user needs to use. After running, the code saves the outputs in a dedicated directory. One output table (in ECSV Format) is provided by transient and by catalog with the list of host galaxy candidates identified and its computed probability of chance alignment. Along with the tables, if enabled by the user, another directory is created to save the plots (one by transients and catalogs) created to facilitate investigations of the associations.

\section{Illustrative Examples}
\label{section:illustrativeexample}

Studying the population of galaxies that produces GRBs and the locations of the GRBs inside their hosts, helps to identify and characterise the GRB progenitor and their environment (e.g. \citep{Bloom2002,Berger2010,Chrimes2018,Fong2022}). But the association between a given GRB and its host galaxy is a very complex issue especially when there is no optical or radio afterglow detection providing a sub-arcsecond localisation. For instance, a large offset between the host galaxy and transient location can originate from the 'kick' velocity imparted to the compact object at the time of birth producing short GRBs (e.g. \citep{Bloom1999,Voss2003}). In this context, the \texttt{Galclaim} software is used to identify, confirm or reject host galaxy candidates allowing to investigate and constrain their properties such as their redshift, stellar population and star formation rate \cite{thesisducoin}.

\section{Impact}
\label{section:impact}

The \texttt{Galclaim} software has already been used to study galaxy association in the case of short GRB transients and has proved its efficiency in identifying host galaxies for such events  \cite{thesisducoin}. The \texttt{Galclaim} software is also used by the GRANDMA collaboration to validate or invalidate transient candidates in several follow-up campaign \cite{GRANDMAkilonova,grandmagrb}. Such tools are very useful to characterise archived transient candidates in large sky survey telescopes such as ZTF \cite{ZTFarchive} or LSST \cite{Ivezic2019}. Given its relatively high processing speed, Galclaim is a suitable tool for real-time classification of LSST transients, which require a drastic filter considering the large numbers involved (see for instance \citep{Moller2021}). The current \texttt{Galclaim} version has typical computation time, enabling all the catalogs, of less that 30s for a given transient position. This computation time is by far dominated by the time needed to make the request in catalogs servers, hence is mainly independent of \texttt{Galclaim} code optimization. This typical computation time allows us to treat about 3000 transient candidates per days. While this is several orders of magnitude bellow the total number of transient candidate that LSST will provide per days, this is compatible with the rate of transient candidate one can get after applying filters dedicated to the search of a given transient source (afterglow, supernova, kilonova...).

\section{Conclusion}
\label{section:conclusion}

The \texttt{Galclaim} software, dedicated to identify association between astrophysical transient sources and host galaxy, is a very useful tool to identify, confirm or reject transient candidates and host galaxy candidates. It's distributed as an open source Python software. It's has already been used by the GRANDMA collaboration and is expected to be widely used in the future for the classification of LSST transients.

\section*{Acknowledgments}
\label{section:acknowledgments}

This project was supported by a research grant from the Ile-de-France Region within the framework of the Domaine d’Intérêt Majeur-Astrophysique et Conditions d’Apparition de la Vie (DIM-ACAV). We acknowledge financial support from the Centre National d’Études Spatiales (CNES). We thank David Corre for his support in this work.





\end{document}